# Decoherence, Correlation, and Entanglement in a Pair of Coupled Quantum Dissipative Oscillators

**A. K. Rajagopal and R.W. Rendell**
Naval Research Laboratory, Washington D. C. 20375-5320

## ABSTRACT

A pair of coupled quantum dissipative oscillators, serving as a model for a nanosystem, is here described by the Lindblad equation. Its dynamic evolution is shown to exhibit the features of decoherence (spatial extent of quantum behavior), correlation (spatial scale over which the system localizes to its physical dimensions), and entanglement (a special quantum feature making its appearance first in such bipartite systems) as a function of the coupling constants of the Lindblad equation. One interesting feature emerging out of this calculation is that the entanglement may exhibit revivals in time. An initially entangled state need not remain so for all time and may exhibit regions of nonentanglement. Interpreting the parameters of the Lindblad theory as environmental features in certain experimental situations, this model calculation gives us clues to possible control of decoherence, correlation, and entanglement. We indicate possible interpretation of the Lindblad parameters as control parameters in more general contexts of some recent experiments.

# I. INTRODUCTION

A prototypical model of many physical systems is often a pair of coupled quantum dissipative oscillators. This model embodies the physical ideas of decoherence, correlation, entanglement, etc. associated with quantum systems. In particular, quantum entanglement shows up for the first time when we consider a two-oscillator system. This is the continuous variable version of the two-qubit system where quantum entanglement raises its head for the first time. The continuum version is of considerable interest both experimentally and theoretically. Only recently, its inseparability criteria were worked out [1, 2]. This is especially pertinent in discussing quantum nanometric systems here by an oscillator-like model, which are usually imbedded in other systems, so that a suitable description of the environmental effects may be described in terms of the parameters of such a model. The purpose of this paper is to exhibit this model in general terms of a Gaussian density matrix containing all the above elements. The "ambiguity function" defined as the Fourier transform in the center-of-mass coordinate of the density matrix is found to lead to solutions of the suitably constructed Lindblad's quantum dissipative oscillators. We use the solution so obtained in estimating the various physical features mentioned above. An important feature of the Lindblad theory is that we can obtain a mixed state from a pure state and vice versa. We restate this property finally, as a conclusion about possible control of decoherence and entanglement in nanometric systems, by varying the coupling constants appearing in the Lindblad equation [3].

We begin by giving a brief account of the density matrix and its two associated functions, the Wigner and ambiguity functions, and their salient properties. A quick survey of the Gaussian forms for these functions and the corresponding interpretations and significance of the coefficients in terms of the concepts of decoherence, correlation, entanglement, and uncertainty relation are also given. The Lindblad equation for the density matrix of a pair of coupled dissipative oscillators is set up and solved using this parameterized Gaussian form. Implications of such analysis to possible practical nanometric systems are indicated.

The Lindblad theory [3] gives a formally exact quantum dynamical equation for the time-dependent density matrix, possessing desirable properties of preserving positivity of the underlying density matrix and including the possibility of passage from pure state to mixed state and vice versa. This theory has recently been applied to study practical applications in condensed matter physics and chemistry. A general framework for dealing with a single dissipative harmonic oscillator in this theory was given by Isar et. al. [4] from which they were able to derive several existing models of dissipation as special cases. In view of its importance and due to lack of methods to solve the Lindblad equation, an action principle was constructed recently [5] as a possible avenue for obtaining approximate solutions. For a discussion of some aspects of decoherence and dissipation using the Lindblad formalism, we may refer to several articles in the book by Giulini et.al. [6], in particular the articles by Joos therein. The solution of the Lindblad equation for the density matrix of a single dissipative oscillator has been studied before using the Gaussian ansatz for the Wigner function and the density matrix [7, 8]. These give rise to complicated coupled equations for the coefficients appearing in the Gaussian ansatz. We present here the solution for a 2-oscillator system in terms of the Gaussian ansatz for the ambiguity function, which yields linear equations for the coefficients, which are then solved in a straightforward way. A similar analysis for the single oscillator case has also been examined recently by us [9].

In the literature, we often find (A) theoretical proposals for future experiments [10, 11, 12] and (B) preliminary experiments [13, 14, 15] on simple coherent systems such as quantum dots, trapped ions, and nuclear spins using magnetic resonance to examine issues of decoherence, entanglement and their control. There is also a recent experimental investigation of controlling the decoherence by coupling to engineered reservoirs [16]. This is done by laser fields, which can change the interaction between a trapped ion and the reservoir. In this paper we discuss some of these in the same exploratory spirit.

In section II, we give the general theoretical framework including the general Lindblad equation for the density matrix as well as the corresponding equation for the ambiguity function. In section III, we introduce the Gaussian ambiguity function and the associated density matrix along with the derivation of physical quantities associated with correlation, decoherence, and entanglement. In section IV, we describe the simplified model of entangled Gaussian due to Simon [1] but in our language. In section V, we construct a dynamical model based on a simplified form for the Lindblad equation given in sec. II, and present and interpret its solution in graphical form using the Gaussian ansatz. In Appendix A, we give the equations for the coefficients in the Gaussian ansatz for the 2-oscillator system in the most general choice of the Lindblad equation. In Appendix B, we give the explicit solution of these equations in a simplified model given in sec.

## II. GENERAL THEORETICAL FRAMEWORK

For simplicity of presentation, we consider the pair of oscillators as a single, two-dimensional system. If $x$ stands for the first system, A, and $y$ for the second, B, we denote them together by a two-dimensional vector $\vec{r} \equiv (x, y)$. We define the time-dependent density matrix in the usual way:

$$\langle \vec{r} | \rho(t) | \vec{r}' \rangle = \langle \vec{r}' | \rho(t) | \vec{r} \rangle^* \text{ (Hermiticity)};$$
$$Tr\rho(t) = \int d^2\vec{r} \langle \vec{r} | \rho(t) | \vec{r} \rangle = 1 \text{ (Trace class); and} \tag{1}$$
$$\int\int d^2\vec{r}\, d^2\vec{r}'\, \phi^*(\vec{r})\langle \vec{r} | \rho(t) | \vec{r}' \rangle \phi(\vec{r}') \geq 0 \text{ (positive semidefiniteness)}.$$

Here * stands for complex conjugate. We define the center of mass and relative coordinates, $\vec{r} + \vec{r}' = 2\vec{R}$, $\vec{r} - \vec{r}' = \mathbf{\vec{r}}$, and define the density matrix in the form

$$\left\langle \vec{R} + \frac{1}{2}\mathbf{\vec{r}} \middle| \rho(t) \middle| \vec{R} - \frac{1}{2}\mathbf{\vec{r}} \right\rangle \equiv \rho(\vec{R}, \mathbf{\vec{r}}; t) \tag{2}$$

Throughout we use units where the Planck constant, $\hbar = 1$. The "ambiguity function" $A(\vec{Q}, \mathbf{\vec{r}}; t)$, is defined as the Fourier transform of the density matrix with respect to $\vec{R}$, and the "Wigner function" $f(\vec{R}, \vec{p}; t)$ as the Fourier transform with respect to $\mathbf{\vec{r}}$, as follows:

$$\rho(\vec{R}, \mathbf{\vec{r}}; t) = \int \frac{d^2\vec{Q}}{(2\pi)^2} e^{-i\vec{Q}\cdot\vec{R}} A(\vec{Q}, \mathbf{\vec{r}}; t)$$
$$= \int \frac{d^2\vec{p}}{(2\pi)^2} e^{-i\vec{p}\cdot\mathbf{\vec{r}}} f(\vec{R}, \vec{p}; t) \tag{3}$$

The properties listed in eq. (1) are reflected as the corresponding properties of the two functions defined above as follows:

$$A^*(\vec{Q},\vec{r};t) = A(-\vec{Q},-\vec{r};t), \text{ and } f^*(\vec{R},\vec{p};t) = f(\vec{R},\vec{p};t) \tag{3a}$$

and the normalization condition

$$A(\vec{Q}=0,\vec{r}=0;t) = 1 = \iint \frac{d^2\vec{R}\,d^2\vec{p}}{(2\pi)^2} f(\vec{R},\vec{p};t) \tag{3b}$$

Also $\quad f(\vec{R},\vec{p};t) = \int d^2\vec{r} \int \frac{d^2\vec{Q}}{(2\pi)^2} e^{-i\vec{Q}\cdot\vec{R}} e^{i\vec{p}\cdot\vec{r}} A(\vec{Q},\vec{r};t) \tag{3c}$

We first observe that there are ten independent covariances (correlations) among the variables of the two systems, which can all be expressed in terms of the various derivatives of the ambiguity function. Here we express these ten covariances of interest that make up the basic uncertainty relations that characterize the system as follows:

$$\begin{aligned}
\langle R_i R_j \rangle &= \iint \frac{d^2\vec{R}\,d^2\vec{p}}{(2\pi)^2} R_i R_j f(\vec{R},\vec{p};t) = -\frac{\partial^2}{\partial Q_i \partial Q_j} A(\vec{Q},\vec{r};t)\bigg|_0 ; \\
\langle p_i p_j \rangle &= \iint \frac{d^2\vec{R}\,d^2\vec{p}}{(2\pi)^2} p_i p_j f(\vec{R},\vec{p};t) = -\frac{\partial^2}{\partial r_i \partial r_j} A(\vec{Q},\vec{r};t)\bigg|_0 ; \text{ and} \\
\langle R_i p_j \rangle &= \iint \frac{d^2\vec{R}\,d^2\vec{p}}{(2\pi)^2} R_i p_j f(\vec{R},\vec{p};t) = \frac{\partial^2}{\partial Q_i \partial r_j} A(\vec{Q},\vec{r};t)\bigg|_0 .
\end{aligned} \tag{4}$$

In the above, i, j go over the two system variables, x, y. Here all the derivatives of the function $A(\vec{Q},\vec{r};t)$ are evaluated at $\vec{Q}=0=\vec{r}$.

The Lindblad equation for the density matrix of a dissipative quantum system is

$$i\partial_t \hat{\rho} = [\hat{H},\hat{\rho}] - \frac{i}{2}\sum_{m,n} h_{nm}\left(\hat{L}_m \hat{L}_n \hat{\rho} + \hat{\rho}\hat{L}_m \hat{L}_n - 2\hat{L}_n \hat{\rho}\hat{L}_m\right) \tag{5}$$

Here $\partial_t = \partial/\partial t$, is the time derivative operator. The first term in the right hand side is the commutator of the Hamiltonian operator of the system $\hat{H}$ representing the usual unitary Hamiltonian evolution, and the second term is the nonunitary evolution governed by a suitably chosen set of Hermitian Lindblad operators $\{\hat{L}_n\}$, and $h_{nm}$ are c-number Hermitian matrx elements to be chosen appropriately to suit the physics of the problem at hand. These properties guarantee the hermiticity of the density matrix and its positivity is assured if the c-number matrix is positive semi-definite. In this paper, we choose for simplicity of presentation, the Hamiltonian to be that of two noninteracting effective oscillators representing the system under consideration:

$$H = \frac{\omega_A}{2}\left(\hat{p}_x^2 + \hat{x}^2\right) + \frac{\omega_B}{2}\left(\hat{p}_y^2 + \hat{y}^2\right) \tag{6}$$

The position and its conjugate momentum operators of each system obey the usual canonical commutation rules and the two systems being independent, the operators belonging to the separate system commute between them. We choose the four Hermitian operators, the position and momentum operators for the two oscillators for the set of $\{\hat{L}_n\}$ operators and the sixteen c-number hermitian coefficients are left unspecified to keep the development general. We also choose for simplicity of presentation, position and momentum variables in dimensionless form so that all the Lindblad parameters have dimensions of energy (recall that we use units with the usual Planck constant is chosen to be unity). The time variable is similarly chosen to be dimensionless, $\tau$, by introducing an energy variable, $\lambda$. They may be chosen later to suit the specific problem at hand at a later stage. Thus,

$$\hat{L}_1 = \hat{x}, \quad \hat{L}_2 = \hat{y}, \quad \hat{L}_3 = \hat{p}_x, \text{ and } \hat{L}_4 = \hat{p}_y. \tag{7}$$

And, we choose for the dissipative part the following most general form:

$$\sum_{m,n} h_{nm}\left(\hat{L}_m\hat{L}_n\hat{\rho} + \hat{\rho}\hat{L}_m\hat{L}_n - 2\hat{L}_n\hat{\rho}\hat{L}_m\right) \equiv \sum_{m,n} h_{nm}\left(\hat{L}_m\hat{L}_n\hat{\rho} + \cdots\right) =$$
$$h_{11}\left(\hat{x}^2\hat{\rho} + \cdots\right) + h_{33}\left(\hat{p}_x^2\hat{\rho} + \cdots\right) + h_{22}\left(\hat{y}^2\hat{\rho} + \cdots\right) + h_{44}\left(\hat{p}_y^2\hat{\rho} + \cdots\right) +$$
$$h_{12}\left(\hat{x}\hat{y}\hat{\rho} + \cdots\right) + h_{12}^*\left(\hat{y}\hat{x}\hat{\rho} + \cdots\right) + h_{13}\left(\hat{x}\hat{p}_x\hat{\rho} + \cdots\right) + h_{13}^*\left(\hat{p}_x\hat{x}\hat{\rho} + \cdots\right) + \tag{8}$$
$$h_{14}\left(\hat{x}\hat{p}_y\hat{\rho} + \cdots\right) + h_{14}^*\left(\hat{p}_y\hat{x}\hat{\rho} + \cdots\right) + h_{23}\left(\hat{y}\hat{p}_x\hat{\rho} + \cdots\right) + h_{23}^*\left(\hat{p}_x\hat{y}\hat{\rho} + \cdots\right) +$$
$$h_{24}\left(\hat{y}\hat{p}_y\hat{\rho} + \cdots\right) + h_{24}^*\left(\hat{p}_y\hat{y}\hat{\rho} + \cdots\right) + h_{34}\left(\hat{p}_x\hat{p}_y\hat{\rho} + \cdots\right) + h_{34}^*\left(\hat{p}_y\hat{p}_x\hat{\rho} + \cdots\right)$$

Introducing the notations, $h_{ij} = h_{ij}^{(r)} + ih_{ij}^{(i)}$, and $h_{ii}$ are real, $\partial_x = \partial/\partial x$, etc. the Lindblad equation in the coordinate representation is found to be (overdot denoting time derivative)

$$i\lambda\partial_\tau\langle\vec{r}_1|\hat{\rho}|\vec{r}_2\rangle = \frac{1}{2}\left\{\begin{array}{l}-\omega_A\left(\partial_{x_1}^2 - \partial_{x_2}^2\right) + \omega_A\left(x_1^2 - x_2^2\right)\\-\omega_B\left(\partial_{y_1}^2 - \partial_{y_2}^2\right) + \omega_B\left(y_1^2 - y_2^2\right)\end{array}\right\}\langle\vec{r}_1|\hat{\rho}|\vec{r}_2\rangle -$$

$$-\frac{i}{2}\left\{\begin{array}{l}h_{11}(x_1 - x_2)^2 - h_{33}\left(\partial_{x_1} + \partial_{x_2}\right)^2 + h_{22}(y_1 - y_2)^2 - h_{44}\left(\partial_{y_1} + \partial_{y_2}\right)^2 +\\ 2h_{12}^{(r)}(x_1 - x_2)(y_1 - y_2) + 2ih_{12}^{(i)}(x_1y_2 - x_2y_1) -\\ -2ih_{13}^{(r)}(x_1 - x_2)\left(\partial_{x_1} + \partial_{x_2}\right) - 2h_{13}^{(i)}\left(1 + x_2\partial_{x_1} + x_1\partial_{x_2}\right) -\\ -2ih_{14}^{(r)}(x_1 - x_2)\left(\partial_{y_1} + \partial_{y_2}\right) - 2h_{14}^{(i)}\left(x_2\partial_{y_1} + x_1\partial_{y_2}\right) -\\ -2ih_{23}^{(r)}(y_1 - y_2)\left(\partial_{x_1} + \partial_{x_2}\right) - 2h_{23}^{(i)}\left(y_2\partial_{x_1} + y_1\partial_{x_2}\right) -\\ -2ih_{24}^{(r)}(y_1 - y_2)\left(\partial_{y_1} + \partial_{y_2}\right) - 2h_{24}^{(i)}\left(1 + y_2\partial_{y_1} + y_1\partial_{y_2}\right) -\\ -2h_{34}^{(r)}\left(\partial_{x_1} + \partial_{x_2}\right)\left(\partial_{y_1} + \partial_{y_2}\right) + 2ih_{34}^{(i)}\left(\partial_{x_1}\partial_{y_2} - \partial_{y_1}\partial_{x_2}\right) -\end{array}\right\}\langle\vec{r}_1|\hat{\rho}|\vec{r}_2\rangle \tag{9}$$

In terms of the center of mass and relative coordinates introduced in eq. (2), we derive the equation obeyed by the ambiguity function:

$$\lambda \partial_\tau A(\vec{Q},\vec{r}) = \{\omega_A(-\mathbf{r_1}\partial_{Q_1} + R_1 r_1) + \omega(-\mathbf{r_2}\partial_{Q_2} + R_2 r_2)\}A(\vec{Q},\vec{r}) - $$

$$-\frac{1}{2}\begin{cases} h_{11}\mathbf{r_1}^2 + h_{33}Q_1^2 + h_{22}\mathbf{r_2}^2 + h_{44}Q_2^2 \\ +2h_{12}^{(r)}\mathbf{r_1 r_2} + 2h_{12}^{(i)}(\mathbf{r_1}\partial_{Q_2} - \mathbf{r_2}\partial_{Q_1}) - \\ -2h_{13}^{(r)}\mathbf{r_1}Q_1 + 2h_{13}^{(i)}(Q_1\partial_{Q_1} + \mathbf{r_1}\partial_{\mathbf{r_1}}) - \\ -2h_{14}^{(r)}\mathbf{r_1}Q_2 + 2h_{14}^{(i)}(Q_2\partial_{Q_1} + \mathbf{r_1}\partial_{\mathbf{r_2}}) - \\ -2h_{23}^{(r)}\mathbf{r_2}Q_1 - 2h_{23}^{(i)}(\mathbf{r_2}\partial_{\mathbf{r_1}} - Q_1\partial_{Q_2}) - \\ -2h_{24}^{(r)}\mathbf{r_2}Q_2 + 2h_{24}^{(i)}(Q_2\partial_{Q_2} + \mathbf{r_2}\partial_{\mathbf{r_2}}) - \\ +2h_{34}^{(r)}Q_1 Q_2 + 2h_{34}^{(i)}(Q_2\partial_{\mathbf{r_1}} - Q_1\partial_{\mathbf{r_2}}) \end{cases} A(\vec{Q},\vec{r})$$

(10)

In the next section we examine the Gaussian structure of the ambiguity function and give in detail the various physical implications of such a function.

### III. AMBIGUITY FUNCTION, DENSITY MATRIX, AND THEIR SIGNIFICANCE

The most general Gaussian form for the density matrix is defined by choosing $A(\vec{Q},\vec{r};t)$ in the following form with time-dependent coefficients (all dimensionless in our notation) with zero mean values $\vec{R}$ and $\vec{p}$:

$$A(\vec{Q},\vec{r};t) = \exp-\frac{1}{2}\left(\mathbf{r_i}A_{ij}(t)\mathbf{r_j} + \mathbf{r_i}B_{ij}(t)Q_j + Q_i B_{ji}(t)\mathbf{r_j} + Q_i C_{ij}(t)Q_j\right)$$

$$= \exp-\frac{1}{2}\left\{(\mathbf{r}^T \ Q^T)\begin{pmatrix} A & B \\ B^T & C \end{pmatrix}\begin{pmatrix} \mathbf{r} \\ Q \end{pmatrix}\right\}.$$

(11)

Here i, j run from 1 to 2, and we use the convention that the repeated indices are summed. The second expression in the above is in terms of a convenient partitioned matrix notation, and the superscript T stands for transposition. In this section, we suppress the time dependence but in the next section when we consider the solution of the Lindblad equation, we exhibit this explicitly. From eqs. (4) and (11), we obtain,

$$\langle R_i R_j \rangle = C_{ij}, \langle p_i p_j \rangle = A_{ij}, \text{ and } \langle R_i p_j \rangle = B_{ji}$$

(12)

and condition (3a) imposes the following requirements on the coefficients in eq. (11), which are seen to be satisfied by virtue of the above identification while condition (3b) is fulfilled by construction:

$A_{ij}$, $C_{ij}$ are real, symmetric and $B_{ij}(\neq B_{ji})$ is real.

$A_{ii}$, $C_{ii} > 0$, $\text{sgn}(B_{ii})$ nonspecific. \hfill (13)

Introducing the matrix notation for the vector denoted now as a column vector and the coefficients A, B, C as matrices, eq. (11) may be expressed in a compact form:

$$A(\vec{Q},\vec{r};t) = \exp-\frac{1}{2}\left(\vec{r}^T \underline{A}(t)\vec{r} + \vec{r}^T \underline{B}(t)\vec{Q} + \vec{Q}^T \underline{B}^T \vec{r} + \vec{Q}^T \underline{C}(t)\vec{Q}\right) \tag{14}$$

Then we deduce the density matrix from eq. (3):

$$\rho(\vec{R},\vec{r}) = \frac{1}{2\pi\sqrt{\det(\underline{C})}} \exp-\frac{1}{2}\left\{(\vec{R}^T,\vec{r}^T)\begin{pmatrix} \underline{C}^{-1} & -i\underline{E} \\ -i\underline{E}^T & (\underline{\alpha}) \end{pmatrix}\begin{pmatrix}\vec{R}\\ \vec{r}\end{pmatrix}\right\} \tag{15}$$

where $\underline{E} = \underline{C}^{-1}\underline{B}^T$, $\underline{\alpha} \equiv \underline{A} - \underline{D} = \underline{A} - \underline{B}\underline{C}^{-1}\underline{B}^T$.

Here $\underline{C}^{-1} = (Det\underline{C})^{-1}\begin{pmatrix} C_{22} & -C_{12} \\ -C_{12} & C_{11} \end{pmatrix}$, and $\underline{B}^T = \begin{pmatrix} B_{11} & B_{21} \\ B_{12} & B_{22} \end{pmatrix}$ \hfill (16)

It is to be noted that the matrix $\underline{D}$ is symmetric upon explicit calculation.

The Wigner function is found to be

$$f_W(\vec{R},\vec{p}) = \frac{1}{\sqrt{\det(\underline{C})}\sqrt{\det(\underline{\alpha})}} \exp-\frac{1}{2}\left\{(\vec{R}^T,\vec{p}^T)\begin{pmatrix} \underline{E}\underline{\alpha}^{-1}\underline{A}\underline{B}^{T-1} & -\underline{E}\underline{\alpha}^{-1} \\ -\underline{\alpha}^{-1}\underline{E}^T & \underline{\alpha}^{-1} \end{pmatrix}\begin{pmatrix}\vec{R}\\ \vec{p}\end{pmatrix}\right\} \tag{17}$$

The reduced density matrices for the two subsystems separately are obtained by the trace operation. We thus obtain the marginal density matrix of system A:

$$\rho_1(R_1,\mathbf{r_1};t) = (2\pi C_{11})^{-1/2}$$
$$\exp-(2C_{11})^{-1}\left(R_1^2 - 2iB_{11}R_1\mathbf{r_1} + (A_{11}C_{11} - B_{11}^2)\mathbf{r_1^2}\right) \tag{18}$$

A similar calculation shows the reduced density matrix of the second system to be:

$$\rho_2(R_2,\mathbf{r_2};t) = (2\pi C_{22})^{-1/2}$$
$$\exp-(2C_{22})^{-1}\left(R_2^2 - 2iB_{22}R_2\mathbf{r_2} + (A_{22}C_{22} - B_{22}^2)\mathbf{r_2^2}\right) \tag{19}$$

It is worth pointing out that these marginal density matrices of the subsystems do not contain remnants from the original two-system density matrix. This aspect becomes even more transparent in subsequent discussion of the uncertainty principle obeyed by the respective correlations of positions and their conjugate momenta.

Following the discussion given by us for the single dissipative oscillator system [9], we deduce the length scales of correlation and decoherence in the subsystems:

$$\langle x^2 \rangle_A = C_{11} = d_A^2(corr);$$
$$\Omega_A^2 = (A_{11} C_{11} - B_{11}^2) = \frac{(1+\xi_A)}{4(1-\xi_A)} \geq 1/4; \quad (20a)$$
$$d_A^2(decoh) = \langle x^2 \rangle_A / 2\Omega_A^2.$$

and

$$\langle y^2 \rangle_B = C_{22} = d_B^2(corr);$$
$$\Omega_B^2 = (A_{22} C_{22} - B_{22}^2) = \frac{(1+\xi_B)}{4(1-\xi_B)} \geq 1/4; \quad (20b)$$
$$d_B^2(decoh) = \langle y^2 \rangle_B / 2\Omega_B^2.$$

$\xi_{A,B}$ are the mixed state parameters of the two single oscillator systems. We now employ a parameterization of a bipartite Gaussian given by Simon [1] and construct the associated ambiguity function and the density matrix of the system.

The equations obeyed by $A_{ij}$, $B_{ij}$, and $C_{ij}$ in eq. (11) when substituted in eq. (10) are given in Appendix A. We observe that all these coefficients are coupled, implying that correlation, decoherence, and entanglement are all dynamically coupled and influence each other. In sec. V we introduce a simplified model to illustrate the main features of these couplings by means of numerical analysis of the solutions.

## IV. A SIMPLE MODEL BASED ON SIMON's WORK

We now consider a canonical parameterization of the two-variable Gaussian density matrix derived from Simon's [1] work which has all the features of entanglement, decoherence, etc. It consists in the following choice of the correlations:

$$\langle x^2 \rangle = a_1, \langle p_x^2 \rangle = b_1, \langle y^2 \rangle = a_2, \langle p_y^2 \rangle = b_2,$$
$$\langle x y \rangle = a_{12}, \text{ and } \langle p_x p_y \rangle = b_{12}. \text{ All others are zero.} \quad (21)$$

Some basic inequalities are obeyed by these quantities following from the Schwarz and Heisenberg inequalities:

$$\text{Schwarz}: \quad a_1 a_2 - a_{12}^2 \equiv K_A \geq 0; \quad b_1 b_2 - b_{12}^2 \equiv K_B \geq 0. \quad (22)$$

$$\text{Heisenberg}: \quad a_1 b_1 \equiv \Omega_A^2 \geq 1/4; \quad a_2 b_2 \equiv \Omega_B^2 \geq 1/4. \quad (23)$$

The above are for the individual oscillator systems. The bipartite Heisenberg inequality and the condition for entanglement derived from Simon's work read as

Heisenberg: $\qquad a_{12} b_{12} \leq 2 K_A K_B - \frac{1}{8}$, (24a)

Entanglement: $\qquad |a_{12} b_{12}| \leq 2 K_A K_B - \frac{1}{8}$, (24b)

Without giving the details, it is straightforward to verify the following expression for the density matrix associated with the Simon model specified by eq. (21):

$$\left\langle \vec{R} + \frac{1}{2}\vec{r} \middle| \rho_S \middle| \vec{R} - \frac{1}{2}\vec{r} \right\rangle = \frac{1}{2\pi (K_A)^{1/2}} \exp -\frac{1}{2} \left\{ \begin{matrix} (x^2 b_1 + 2xy b_{12} + y^2 b_2) + \\ K_A^{-1}(X^2 a_2 - 2XY a_{12} + Y^2 a_1) \end{matrix} \right\} \qquad (25)$$

Here $x = (x_1 - x_2)$, $y = (y_1 - y_2)$, $X = \frac{1}{2}(x_1 + x_2)$, and $Y = \frac{1}{2}(y_1 + y_2)$.

From this we have the reduced density matrices of A and B subsystems which are found to be both mixed state density matrices:

$$\langle x_1 | \rho_{S,A} | x_2 \rangle = \frac{1}{(2\pi a_1)^{1/2}} \exp -\frac{1}{2}\left\{ \left(b_1 + \frac{1}{4a_1}\right)x_1^2 + \left(b_1 + \frac{1}{4a_1}\right)x_2^2 - 2 x_1 x_2 \left(b_1 - \frac{1}{4a_1}\right) \right\} \qquad (26)$$

$$\langle y_1 | \rho_{S,B} | y_2 \rangle = \frac{1}{(2\pi b_2)^{1/2}} \exp -\frac{1}{2}\left\{ \left(b_2 + \frac{1}{4a_2}\right)y_1^2 + \left(b_2 + \frac{1}{4a_2}\right)y_2^2 - 2 y_1 y_2 \left(b_2 - \frac{1}{4a_2}\right) \right\} \qquad (27)$$

The expressions for eqs. (26, 27) with $x_1 = x_2$ and $y_1 = y_2$ respectively lead to the identification of correlation lengths while those with $x_1 = -x_2$ and $y_1 = -y_2$ respectively into the identification of decoherence lengths in the subsystems:

$$\begin{aligned} d_{S,A}^2(corr) &= a_1; \quad d_{S,A}^2(decoh) = 1/4b_1; \\ d_{S,B}^2(corr) &= a_2; \quad d_{S,B}^2(decoh) = 1/4b_2. \end{aligned} \qquad (28)$$

The mixed state lengths in these subsystems are identified to be the coefficients of the products $x_1 x_2$ and $y_1 y_2$ respectively in eqs. (26, 27):

$$d_{S,A}^2(mix) = 4a_1 / (4\Omega_A^2 - 1) \text{ and } d_{S,B}^2(mix) = 4a_2 / (4\Omega_B^2 - 1) \qquad (29)$$

Similar analysis of the composite system density matrix given by eq. (25) lead to correlation and decoherence lengths:

$$d_{S,AB}^2(A-corr) = K_A/a_2 = d_{S,A}^2(corr) - a_{12}^2/a_2; \quad d_{S,AB}^2(A-decoh) = 1/4b_1 = d_A^2(decoh);$$
$$d_{S,AB}^2(B-corr) = K_A/a_1 = d_{S,B}^2(corr) - a_{12}^2/a_1; \quad d_{S,AB}^2(B-decoh) = 1/4b_2 = d_B^2(decoh).$$
(30)

The coefficients of the products xy and XY in eq. (25) indicate the entanglement features in the composite system, which we here define as entanglement lengths:

$$E_{S,AB}^2 = 1/b_{12} \text{ and } \tilde{E}_{S,AB}^2 = K_A/a_{12}.$$
(31)

Eq. (24) representing the Heisenberg inequality for the bipartite system may be written then in the form

$$E_{S,AB}^{-2}\tilde{E}_{S,AB}^{-2} \leq \left(\frac{1}{4K_A} + 2K_B\right).$$
(32)

In the next section, we develop a model of the two-oscillator system based on the above-simplified parameterization of the Simon model as the given input at initial time in solving the Lindbad equation. This will exhibit how the Lindblad parameters can influence the entanglement features of the bipartite system.

### V. A DYNAMICAL MODEL – SOLUTION OF LINDBLAD EQUATION

The dynamical model described here serves to illustrate how the solution of the Lindblad equation exhibits time-evolution in the initially specified correlation functions given by the Simon model. This model indicates how one may control the parameters specifying the decoherence and entanglement by suitable choice of the interactions introduced in the Lindblad equation. In the general eqs. (A1-A10), we define our model by keeping only the following Lindblad interaction constants and all others are set equal to zero;

$$h_{ii}, i = 1, 2, 3, 4, \text{ real part of } h_{12}, \text{ and both real and imaginary parts of } h_{13}, h_{24}.$$
(33)

The $h_{ii}$'s and the real parts of $h_{13}, h_{24}$ serve as driving forces in the two systems, whereas the imaginary parts of $h_{13}, h_{24}$ give rise to damping of the two oscillators. And, $h_{12}^{(r)}$ serves as the driving force for the entanglement in the system. This choice of the Lindblad parameters simplify the coupled equations in eq. (12) in such a way that the two oscillators A and B are not coupled to each other but are governed by their own individual parameters. Thus their individuals decoherence and correlation features are preserved even in this simple model. Also, the entanglement features appear here as four coupled equations for the cross correlation functions with their own friction forces. The following set of coupled, linear equations describe this simplified model.

$$\lambda \partial_\tau A_{11} = -2\omega_A B_{11} - 2h_{13}^{(i)} A_{11} + h_{11},$$
$$\lambda \partial_\tau B_{11} = \omega_A A_{11} - \omega_A C_{11} - 2h_{13}^{(i)} B_{11} - h_{13}^{(r)},$$ (34)
$$\lambda \partial_\tau C_{11} = 2\omega_A B_{11} - 2h_{13}^{(i)} C_{11} + h_{33}.$$

$$\lambda \partial_\tau A_{22} = -2\omega_B B_{22} - 2h_{24}^{(i)} A_{22} + h_{22},$$
$$\lambda \partial_\tau B_{22} = \omega_B A_{22} - \omega_B C_{22} - 2h_{24}^{(i)} B_{22} - h_{24}^{(r)},$$ (35)
$$\lambda \partial_\tau C_{22} = 2\omega_B B_{22} - 2h_{24}^{(i)} C_{22} + h_{44};$$

$$\lambda \partial_\tau A_{12} = -\omega_A B_{21} - \omega_B B_{12} - \left(h_{13}^{(i)} + h_{24}^{(i)}\right) A_{12} + h_{12}^{(r)},$$
$$\lambda \partial_\tau B_{12} = \omega_B A_{12} - \omega_A C_{12} - \left(h_{13}^{(i)} + h_{24}^{(i)}\right) B_{12},$$
$$\lambda \partial_\tau B_{21} = \omega_A A_{12} - \omega_B C_{12} - \left(h_{13}^{(i)} + h_{24}^{(i)}\right) B_{21},$$ (36)
$$\lambda \partial_\tau C_{12} = \omega_A B_{12} + \omega_B B_{21} - \left(h_{13}^{(i)} + h_{24}^{(i)}\right) C_{12}.$$

Eqs. (34, 35) are the respective equations for the oscillators A, B respectively and are the same as the ones solved in the RWR-AKR [9] paper for a single dissipative oscillator. Eq. (36) on the other hand, are the equations coupling the two oscillators, representing entanglement. These being coupled first order differential equations in time, we specify the initial conditions as in the Simon model:

$$A_{11}(\tau=0) = b_1, B_{11}(\tau=0) = 0, C_{11}(\tau=0) = a_1,$$
$$A_{22}(\tau=0) = b_2, B_{22}(\tau=0) = 0, C_{22}(\tau=0) = a_2,$$ (37)
$$A_{12}(\tau=0) = b_{12}, B_{12}(\tau=0) = 0, B_{21}(\tau=0) = 0, C_{12}(\tau=0) = a_{12}.$$

These equations are solved by the method of Laplace transformation incorporating the initial conditions given by eq. (37). The time dependencies of the coefficients $A_{12}, B_{12}$ given by eqs. (36) are of interest to us as they represent the evolution of entanglement of the two oscillators. In Appendix B, we give the exact analytical solutions of these equations. The numerical display of these results will be described presently. We should remark here that this simple model does not compromise the general features of the system, as will be evident from the foregoing discussion.

It may be worth noting that the above initial conditions and their time evolutions governed by eqs. (34, 35, 36) may be expressed neatly as the evolution of the covariance matrices as follows.

$$A_S(\tau=0) = \begin{pmatrix} a_1 & 0 \\ 0 & b_1 \end{pmatrix} \to A_S(\tau) = \begin{pmatrix} C_{11} & B_{11} \\ B_{11} & A_{11} \end{pmatrix}.$$ (38)

$$B_S(\tau=0) = \begin{pmatrix} a_2 & 0 \\ 0 & b_2 \end{pmatrix} \to B_S(\tau) = \begin{pmatrix} C_{22} & B_{22} \\ B_{22} & A_{22} \end{pmatrix}.$$ (39)

$$C_S(\tau = 0) = \begin{pmatrix} a_{12} & 0 \\ 0 & b_{12} \end{pmatrix} \to C_S(\tau) = \begin{pmatrix} C_{12} & B_{12} \\ B_{21} & A_{12} \end{pmatrix}. \tag{40}$$

From Appendix B, we note how these evolutions come about explicitly.

In this form the Simon inequalities given in eqs. (24a, b) are now written in the form

Heisenberg:
$$(\det A_S)(\det B_S) + \left(\frac{1}{4} - \det C_S\right)^2 - tr\left(A_S J C_S J B_S J C_S^T J\right) \geq \frac{1}{4}(\det A_S + \det B_S), \tag{41a}$$

Entanglement:
$$(\det A_S)(\det B_S) + \left(\frac{1}{4} - |\det C_S|\right)^2 - tr\left(A_S J C_S J B_S J C_S^T J\right) \geq \frac{1}{4}(\det A_S + \det B_S), \tag{41b}$$

Here $J = \begin{pmatrix} 0 & 1 \\ -1 & 0 \end{pmatrix}$ and $C_S^T$ is the transposed matrix of $C_S$. The Schwarz and Heisenberg inequalities for A and B systems are

Schwarz: $C_{11}C_{22} - C_{12}^2 \equiv K_A \geq 0; \quad A_{11}A_{22} - A_{12}^2 \equiv K_B \geq 0.$ \hfill (42)

Heisenberg: $\det A_S \equiv \Omega_A^2 = A_{11}C_{11} - B_{11}^2 \geq 1/4; \quad \det B_S \equiv \Omega_B^2 = A_{22}C_{22} - B_{22}^2 \geq 1/4.$ (43)

As is pointed out by Simon [1], it is sufficient to examine the sign of the $\det C_S$ to determine whether one has entanglement (if the sign is negative) or not (if the sign is zero or positive). In the numerical work presented here for the simplified model worked out in Appendix B, we deduce analytically an important result for asymptotically large times. In fact, we have

$$\det C_S(\tau = \infty) = \left(\frac{h_{12}^{(r)}}{\omega_A\left[\Gamma^2 + (1+r)^2\right]\left[\Gamma^2 + (1-r)^2\right]}\right)^2 \{(r-1)\Gamma^4 + \Gamma^2(r^3 - r^2 + 2r - 2) - (r^2 - r + 1)\} \tag{44}$$

This result, it should be noted is independent of the initial conditions. For finite times, however, a more complicated calculation needs to be made, which will be presented graphically in this paper. Corresponding to the graphical presentations, we give here the results for two typical cases.

Case (1): r=1 (equivalent oscillators) we find from eq. (44)

$$\det C_S(\tau = \infty) < 0. \tag{45a}$$

Case (2): $r \neq 1$ (inequivalent oscillators)
In this case, depending on the values of r and $\Gamma$ we may have

$$\det C_S(\tau = \infty) > 0 \quad \text{and} \quad \det C_S(\tau = \infty) < 0. \tag{45b}$$

These results show that the irrespective of the initial conditions, for suitable choice of Lindblad parameters we can get unentangled or entangled state.

We now present detailed calculations of the time evolution of decoherence and entanglement with given initial conditions. We choose $\lambda = \omega_A$ and use position and momentum variables in dimensionless form so that all Lindblad parameters have dimensions of energy and the dimensionless time variable is $\tau = \omega_A t$. The choice of the parameters for the initial conditions must be consistent with eqs. (22, 23, 24a,b). We choose here two special choices for purposes of illustration with minimum uncertainty values $a_1 = b_1 = 1/2$, $a_2 = b_2 = 1/2$:

(A) Initially unentangled state with $a_{12} = 0 = b_{12}$,

(B) Initially entangled state with $a_{12} = 1/2 = -b_{12}$.

From eqs. (47a,b), we deduce that one may get an unentangled or entangled state in both these situations for suitable choice of the Lindblad parameters, which are chosen to preserve the positive signs of the mean square displacements and momenta of the two oscillators. We focus on the case of inequivalent oscillators by choosing their frequencies to be different (the frequency of oscillator B is here chosen to be three times faster than that of A) but all other parameters were chosen to be the same for convenient presentation of the results. Their values are given in the caption of Fig.1, and are kept the same in calculating all other system characteristics.

In Fig.1, we display the mean square momentum of the two subsystems as a function of the dimensionless time $\tau$. Since we assumed the two oscillators to be in their minimum uncertainty states, they both begin at the same value (0.5) initially and evolve according to the solution given in Appendix B, eq. (B3) and its counterpart. They approach their respective asymptotes for large times, the B-oscillator approaching it much earlier than the A. This is due to the fact that the decay constants are chosen to be the same for both the systems.

In Fig.2, we present the subsystem decoherence lengths defined by eqs. (20a,b) and the solutions given in Appendix B. It is interesting to note that as in the case of momenta in Fig.1, the decoherence length crosses the A system value for times of about 0.6, and approach their asymptotic values for large times (for times larger than 5 on this Figure). This is because the decoherence length is a ratio of similarly decaying quantities.

Fig.3 represents the oscillator pair correlations involving positions and momenta of the two systems. These are important in determining the dynamic evolution of entanglement. In Fig.3a, the initial values for these are chosen to be zero, which corresponds to Case (A) above when the system is initially unentangled. Fig. 3b, on the other hand, is for the case when they are initially entangled. They both oscillate in approximately opposite phase, the second case exhibiting more oscillations than the first. They change their signs a few times before reaching, albeit slowly, their respective asymptotic values.

Finally, Fig.4, displays the time evolution of the determinant constructed from the solutions given by eqs. (B4-7) of Appendix B. This is a signature of "entanglement" of the systems A and B. The curve (a) is for the initially unentangled case whereas curve (B) is for the initially entangled case. They both show oscillations about zero values, exhibiting "revival" of entanglement as time progresses. This also clearly shows that the entanglement property changes over time, a feature worth emphasizing. What was initially unentangled may become entangled some time later and vice versa and this characteristic may change several times over a period of time. This implies that in actual experiment, such an oscillation in entanglement may provide windows where such properties are either to be preferred or avoided. It should be remarked that these features of the simplified model are retained more or less in the forms presented here when one considers more general equations given in Appendix A. The only point to be noted is that the decoherence and entanglement influence each other in this more general setting, which was not the case in the model discussed here.

## VI. CONCLUDING REMARKS

One of the important consequences of the Lindblad theory is that one can obtain a mixed state from an initial pure state and vice versa. In this paper, in particular, we demonstrate another aspect of this feature by showing that we can obtain an entangled state from an initially unentangled state (pure or mixed). This feature allows us to reinterpret it as a manipulation of the entanglement by means of the parameters of the theory, which in turn is a manifestation of the environment or other elements of the system. Manipulation of decoherence is also possible as is clear from our example. Interpreting the Lindblad parameters as the parameters associated with the environment, we note that the model calculation given here implies that important features of decoherence and entanglement may be manipulated by suitable change in the environment. In the experimental situations presented in refs. [13-15], for example, simple coherent systems such as quantum dots, trapped ions, and nuclear spins are studied for realizing these features and their possible control. There is also a recent experimental investigation [16] of controlling the decoherence of trapped ions by laser fields which can change the interaction between the trapped ion and the reservoir. The same technique may possibly be employed to investigate the control of entanglement. We hope to investigate possible determinations of the Lindblad parameters in terms of interactions between the environment Hamiltonian and the system of interest, thus providing a microscopic picture of such phenomena in realistic situations.
-

## ACKNOWLEDGEMENTS

Both the authors are supported in part by the Office of Naval Research. We also thank Dr. Peter Reynolds of the Office Naval Research for supporting this research. We dedicate this paper to the memory of Professor Chanchal Majumdar who lived by the principle that work is its own reward.

## APPENDIX A: DYNAMICAL EQUATIONS FOR THE MATRIX ELEMENTS

The equations for the coefficients in the Gaussian ansatz, eq. (11) are obtained when substituted in eq. (10). Here we arrange them in three sets; the first corresponds to the oscillator A, the second to the oscillator B, while the third set corresponds to the interaction between the two oscillators.

$$\lambda \partial_\tau A_{11} = -2\omega_A B_{11} - 2h_{13}^{(i)} A_{11} - 2h_{12}^{(i)} B_{12} - 2h_{14}^{(i)} A_{12} + h_{11},$$
$$\lambda \partial_\tau B_{11} = -\omega_A A_{11} - \omega_A C_{11} - 2h_{13}^{(i)} B_{11} + h_{34}^{(i)} A_{12} - 2h_{23}^{(i)} B_{12} - h_{14}^{(i)} B_{21} - h_{12}^{(i)} C_{12} - h_{13}^{(r)},$$
$$\lambda \partial_\tau C_{11} = 2\omega_A B_{11} - 2h_{13}^{(i)} C_{11} + 2h_{34}^{(i)} B_{21} - 2h_{23}^{(i)} C_{12} + h_{33}.$$

(A1, 2, 3)

$$\lambda \partial_\tau A_{22} = 2\omega_B B_{22} - 2h_{24}^{(i)} A_{22} + 2h_{23}^{(i)} A_{12} + 2h_{12}^{(i)} B_{21} + h_{22},$$
$$\lambda \partial_\tau B_{22} = \omega_B A_{22} - \omega_B C_{22} - 2h_{24}^{(i)} B_{22} + h_{34}^{(i)} A_{12} + h_{23}^{(i)} B_{12} - h_{14}^{(i)} B_{21} + h_{12}^{(i)} C_{12} - h_{24}^{(r)},$$
$$\lambda \partial_\tau C_{22} = 2\omega_B B_{22} - 2h_{24}^{(i)} C_{22} - 2h_{34}^{(i)} B_{12} - 2h_{14}^{(i)} C_{12} + h_{44};$$

(A4, 5, 6)

$$\lambda \partial_\tau A_{12} = -\omega_A B_{21} - \omega_B B_{12} - \left(h_{13}^{(i)} + h_{24}^{(i)}\right) A_{12} + h_{23}^{(i)} A_{11} - h_{14}^{(i)} A_{22} + h_{12}^{(i)} B_{11} - h_{12}^{(i)} B_{22} + h_{12}^{(r)},$$
$$\lambda \partial_\tau B_{12} = \omega_B A_{12} - \omega_A C_{12} - \left(h_{13}^{(i)} + h_{24}^{(i)}\right) B_{12} - h_{34}^{(i)} A_{11} - h_{14}^{(i)} B_{11} - h_{14}^{(i)} B_{22} - h_{12}^{(i)} C_{22} - h_{14}^{(r)},$$
$$\lambda \partial_\tau B_{21} = \omega_A A_{12} - \omega_B C_{12} - \left(h_{13}^{(i)} + h_{24}^{(i)}\right) B_{21} - h_{34}^{(i)} A_{22} + h_{23}^{(i)} B_{11} - h_{23}^{(i)} B_{22} - h_{12}^{(i)} C_{11} + h_{23}^{(r)},$$
$$\lambda \partial_\tau C_{12} = \omega_A B_{12} + \omega_B B_{21} - \left(h_{13}^{(i)} + h_{24}^{(i)}\right) C_{12} - h_{34}^{(i)} B_{11} + h_{34}^{(i)} B_{11} - h_{14}^{(i)} C_{11} - h_{23}^{(i)} C_{22} + h_{34}^{(r)}.$$

(A7, 8, 9, 10)

We note that these equations are coupled to each other in an interesting way. The equations for the A and B oscillators are coupled to each other via their interactions introduced in the Lindblad evolution. This implies that in this model, the correlations and decoherence in the two oscillators are influenced by the entanglement between the two due to the dissipative processes contained in the Lindblad formulation. The stationary solutions of these equations are obtained by setting to zero all the time derivatives in the left sides of these equations. They are also the solutions approached for asymptotically large times. In a simplified model given in the text, we consider a decoupled set of equations, which do not compromise the final results but serves the purpose of illustrating how these influences come about and how one may control these important features of the quantum oscillator pair. In the Appendix B, explicit solutions of the equations are presented.

## APPENDIX B: SOLUTIONS OF EQUATIONS (34, 35, and 36)

The solution for the A-oscillator equations (eq.34) is:

$$A_{11}(\tau) = \frac{1}{2} b_1 e^{-\tau\left(\frac{\omega_A}{\lambda}\right)\Gamma_A}\left(1+\cos 2\tau\left(\frac{\omega_A}{\lambda}\right)\right) + \frac{1}{2} a_1 e^{-\tau\left(\frac{\omega_A}{\lambda}\right)\Gamma_A}\left(1-\cos 2\tau\left(\frac{\omega_A}{\lambda}\right)\right)$$

$$+\frac{(h_{11}/\omega_A)}{\Gamma_A(\Gamma_A^2+4)}\left\{\begin{array}{l}\Gamma_A^2\left[1-\dfrac{e^{-\tau\left(\frac{\omega_A}{\lambda}\right)\Gamma_A}}{2}\left(1+\cos 2\tau\left(\frac{\omega_A}{\lambda}\right)\right)\right] \\ +2\left(1-e^{-\tau\left(\frac{\omega_A}{\lambda}\right)\Gamma_A}\right)+\Gamma_A e^{-\tau\left(\frac{\omega_A}{\lambda}\right)\Gamma_A}\sin 2\tau\left(\frac{\omega_A}{\lambda}\right)\end{array}\right\}$$

$$+\frac{(h_{33}/\omega_A)}{2\Gamma_A(\Gamma_A^2+4)}\left\{\begin{array}{l}-\Gamma_A^2 e^{-\tau\left(\frac{\omega_A}{\lambda}\right)\Gamma_A}\left(1-\cos 2\tau\left(\frac{\omega_A}{\lambda}\right)\right) \\ +4\left(1-e^{-\tau\left(\frac{\omega_A}{\lambda}\right)\Gamma_A}\right)-2\Gamma_A e^{-\tau\left(\frac{\omega_A}{\lambda}\right)\Gamma_A}\sin 2\tau\left(\frac{\omega_A}{\lambda}\right)\end{array}\right\}$$

$$+\frac{(h_{13}^{(r)}/\omega_A)}{(\Gamma_A^2+4)}\left\{2\left(1-e^{-\tau\left(\frac{\omega_A}{\lambda}\right)\Gamma_A}\cos 2\tau\left(\frac{\omega_A}{\lambda}\right)\right)-\Gamma_A e^{-\tau\left(\frac{\omega_A}{\lambda}\right)\Gamma_A}\sin 2\tau\left(\frac{\omega_A}{\lambda}\right)\right\} \quad \text{(B1)}$$

$$B_{11}(\tau) = \frac{(b_1-a_1)}{2} e^{-\tau\left(\frac{\omega_A}{\lambda}\right)\Gamma_A}\sin 2\tau\left(\frac{\omega_A}{\lambda}\right)$$

$$+\frac{((h_{11}-h_{33})/\omega_A)}{(\Gamma_A^2+4)}\left\{1-e^{-\tau\left(\frac{\omega_A}{\lambda}\right)\Gamma_A}\left(\cos 2\tau\left(\frac{\omega_A}{\lambda}\right)+\frac{1}{2}\Gamma_A \sin 2\tau\left(\frac{\omega_A}{\lambda}\right)\right)\right\}$$

$$-\frac{(h_{13}^{(r)}/\omega_A)}{(\Gamma_A^2+4)}\left\{\Gamma_A\left(1-e^{-\tau\left(\frac{\omega_A}{\lambda}\right)\Gamma_A}\cos 2\tau\left(\frac{\omega_A}{\lambda}\right)\right)-2e^{-\tau\left(\frac{\omega_A}{\lambda}\right)\Gamma_A}\sin 2\tau\left(\frac{\omega_A}{\lambda}\right)\right\} \quad \text{(B2}$$

$$C_{11}(\tau) = \frac{1}{2} a_1 e^{-\tau\left(\frac{\omega_A}{\lambda}\right)\Gamma_A}\left(1+\cos 2\tau\left(\frac{\omega_A}{\lambda}\right)\right) + \frac{1}{2} b_1 e^{-\tau\left(\frac{\omega_A}{\lambda}\right)\Gamma_A}\left(1-\cos 2\tau\left(\frac{\omega_A}{\lambda}\right)\right)$$

$$+\frac{(h_{11}/\omega_A)}{2\Gamma_A(\Gamma_A^2+4)}\left\{\begin{array}{l}-\Gamma_A^2 e^{-\tau\left(\frac{\omega_A}{\lambda}\right)\Gamma_A}\left(1-\cos 2\tau\left(\frac{\omega_A}{\lambda}\right)\right)\\+4\left(1-e^{-\tau\left(\frac{\omega_A}{\lambda}\right)\Gamma_A}\right)-2\Gamma_A e^{-\tau\left(\frac{\omega_A}{\lambda}\right)\Gamma_A}\sin 2\tau\left(\frac{\omega_A}{\lambda}\right)\end{array}\right\}$$

$$+\frac{(h_{33}/\omega_A)}{2\Gamma_A(\Gamma_A^2+4)}\left\{\begin{array}{l}\Gamma_A^2\left(1-e^{-\tau\left(\frac{\omega_A}{\lambda}\right)\Gamma_A}\cos 2\tau\left(\frac{\omega_A}{\lambda}\right)\right)\\+(\Gamma_A^2+4)\left(1-e^{-\tau\left(\frac{\omega_A}{\lambda}\right)\Gamma_A}\right)+2\Gamma_A e^{-\tau\left(\frac{\omega_A}{\lambda}\right)\Gamma_A}\sin 2\tau\left(\frac{\omega_A}{\lambda}\right)\end{array}\right\}$$

$$+\frac{(h_{13}^{(r)}/\omega_A)}{(\Gamma_A^2+4)}\left\{-2\left(1-e^{-\tau\left(\frac{\omega_A}{\lambda}\right)\Gamma_A}\cos 2\tau\left(\frac{\omega_A}{\lambda}\right)\right)+\Gamma_A e^{-\tau\left(\frac{\omega_A}{\lambda}\right)\Gamma_A}\sin 2\tau\left(\frac{\omega_A}{\lambda}\right)\right\} \quad (B3)$$

In the above expressions we have set $\Gamma_A = 2h_{13}^{(i)}/\omega_A$. The solution for the B-oscillator (eq.37) is obtained by the substitutions $1 \to 2$, $3 \to 4$, and $A \to B$ in the above expressions.

We now give the solution to eq. (36). Here we set $\Gamma = (\Gamma_A + r\Gamma_B)/2$ and $r = \omega_B/\omega_A$:

$$A_{12}(\tau) = b_{12} e^{-\tau(\omega_A/\lambda)\Gamma}\cos\tau(\omega_A/\lambda)\cos\tau(\omega_B/\lambda)$$
$$+ a_{12} e^{-\tau(\omega_A/\lambda)\Gamma}\sin\tau(\omega_A/\lambda)\sin\tau(\omega_B/\lambda)$$
$$+\left(\frac{h_{12}^{(r)}}{\omega_A}\right)\Gamma\frac{(\Gamma^2+1+r^2)}{[\Gamma^2+(1+r)^2][\Gamma^2+(1-r)^2]}$$
$$-\left(\frac{h_{12}^{(r)}}{\omega_A}\right)\frac{\Gamma e^{-\tau(\omega_A/\lambda)\Gamma}}{2}\left\{\frac{\cos\tau(\omega_A+\omega_B)/\lambda}{\Gamma^2+(1+r)^2}+\frac{\cos\tau(\omega_A-\omega_B)/\lambda}{\Gamma^2+(1-r)^2}\right\}$$
$$+\left(\frac{h_{12}^{(r)}}{\omega_A}\right)\frac{e^{-\tau(\omega_A/\lambda)\Gamma}}{2}\left\{\frac{(1+r)\sin\tau(\omega_A+\omega_B)/\lambda}{\Gamma^2+(1+r)^2}-\frac{(1-r)\sin\tau(\omega_A-\omega_B)/\lambda}{\Gamma^2+(1-r)^2}\right\} \quad (B4)$$

$$B_{12}(\tau) = \frac{b_{12}e^{-\tau(\omega_A/\lambda)\Gamma}}{4}\left[\sin\tau(\omega_A(1+r)/\lambda)+\sin\tau(\omega_A(1-r)/\lambda)\right]$$
$$-\frac{a_{12}e^{-\tau(\omega_A/\lambda)\Gamma}}{2}\left[\frac{3}{(1+r)}\sin\tau(\omega_A(1+r)/\lambda)+\frac{1}{(1-r)}\sin\tau(\omega_A(1-r)/\lambda)\right]$$

$$+\left(\frac{h_{12}^{(r)}}{\omega_A}\right)\frac{(\Gamma^2+1+r^2-r)}{\left[\Gamma^2+(1+r)^2\right]\left[\Gamma^2+(1-r)^2\right]}$$

$$-\left(\frac{h_{12}^{(r)}}{\omega_A}\right)\frac{3e^{-\tau(\omega_A/\lambda)\Gamma}}{4(1+r)\left[\Gamma^2+(1+r)^2\right]}\{\Gamma\sin\tau\omega_A(1+r)/\lambda+(1+r)\cos\tau\omega_A(1+r)/\lambda\}$$

$$-\left(\frac{h_{12}^{(r)}}{\omega_A}\right)\frac{e^{-\tau(\omega_A/\lambda)\Gamma}}{4(1-r)\left[\Gamma^2+(1-r)^2\right]}\{\Gamma\sin\tau\omega_A(1-r)/\lambda+(1-r)\cos\tau\omega_A(1-r)/\lambda\}$$

(B5)

$$B_{21}(\tau)=\frac{b_{12}e^{-\tau(\omega_A/\lambda)\Gamma}}{4(1+r)}\left[\frac{(2+r)}{(1+r)}\sin\tau(\omega_A(1+r)/\lambda)+\frac{(2-r)}{(1-r)}\sin\tau(\omega_A(1-r)/\lambda)\right]$$

$$-\frac{a_{12}e^{-\tau(\omega_A/\lambda)\Gamma}}{2}\left[\sin\tau(\omega_A(1+r)/\lambda)+\sin\tau(\omega_A(1-r)/\lambda)\right]$$

$$+\left(\frac{h_{12}^{(r)}}{\omega_A}\right)\frac{(\Gamma^2+1)}{\left[\Gamma^2+(1+r)^2\right]\left[\Gamma^2+(1-r)^2\right]}$$

$$-\left(\frac{h_{12}^{(r)}}{\omega_A}\right)\frac{(2+r)e^{-\tau(\omega_A/\lambda)\Gamma}}{4(1+r)\left[\Gamma^2+(1+r)^2\right]}\{-\Gamma\sin\tau\omega_A(1+r)/\lambda+(1+r)\cos\tau\omega_A(1+r)/\lambda\}$$

$$-\left(\frac{h_{12}^{(r)}}{\omega_A}\right)\frac{(2-r)e^{-\tau(\omega_A/\lambda)\Gamma}}{4(1-r)\left[\Gamma^2+(1-r)^2\right]}\{-\Gamma\sin\tau\omega_A(1-r)/\lambda+(1-r)\cos\tau\omega_A(1-r)/\lambda\}$$

(B6)

$$C_{12}(\tau)=a_{12}e^{-\tau(\omega_A/\lambda)\Gamma}\cos\tau(\omega_A/\lambda)\cos\tau(\omega_B/\lambda)$$

$$+b_{12}e^{-\tau(\omega_A/\lambda)\Gamma}\sin\tau(\omega_A/\lambda)\sin\tau(\omega_B/\lambda)$$

$$+\left(\frac{h_{12}^{(r)}}{\omega_A}\right)\Gamma\frac{r}{\left[\Gamma^2+(1+r)^2\right]\left[\Gamma^2+(1-r)^2\right]}$$

$$+\left(\frac{h_{12}^{(r)}}{\omega_A}\right)\frac{e^{-\tau(\omega_A/\lambda)\Gamma}}{4\left[\Gamma^2+(1+r)^2\right]}\{\Gamma\cos\tau\omega_A(1+r)/\lambda-(1+r)\sin\tau\omega_A(1+r)/\lambda\}$$

$$-\left(\frac{h_{12}^{(r)}}{\omega_A}\right)\frac{e^{-\tau(\omega_A/\lambda)\Gamma}}{4[\Gamma^2+(1-r)^2]}\{\Gamma\cos\tau\omega_A(1-r)/\lambda-(1-r)\sin\tau\omega_A(1-r)/\lambda\}$$

(B7)

These expressions are numerically evaluated for a certain choice of the parameters and are displayed in graphical form in the figures. Their significance is then elucidated in terms of some of the experimental situations that are being examined which were mentioned in the Introduction.

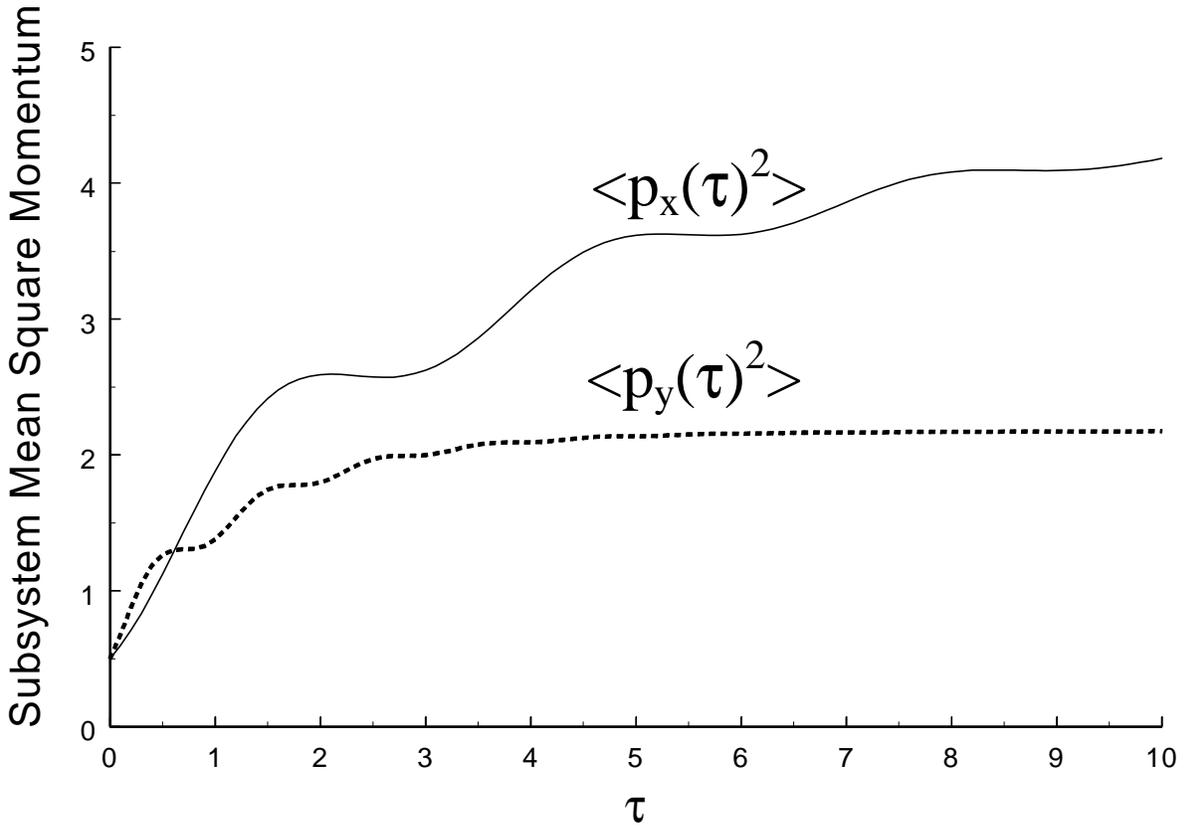

Figure 1. Subsystem mean square momentum $\langle p_x^2 \rangle$ for oscillator A and $\langle p_y^2 \rangle$ for oscillator B in dimensionless units, using the solutions in Appendix B with $\lambda = \omega_A$ and $\Gamma_A = \Gamma_B = 0.25$, $r = \omega_B/\omega_A = 3$, $h_{11} = h_{33} = h_{13}^{(r)} = 1$, $h_{22} = 2$, $h_{44} = 4$, and $a_1 = b_1 = a_2 = b_2 = 0.5$.

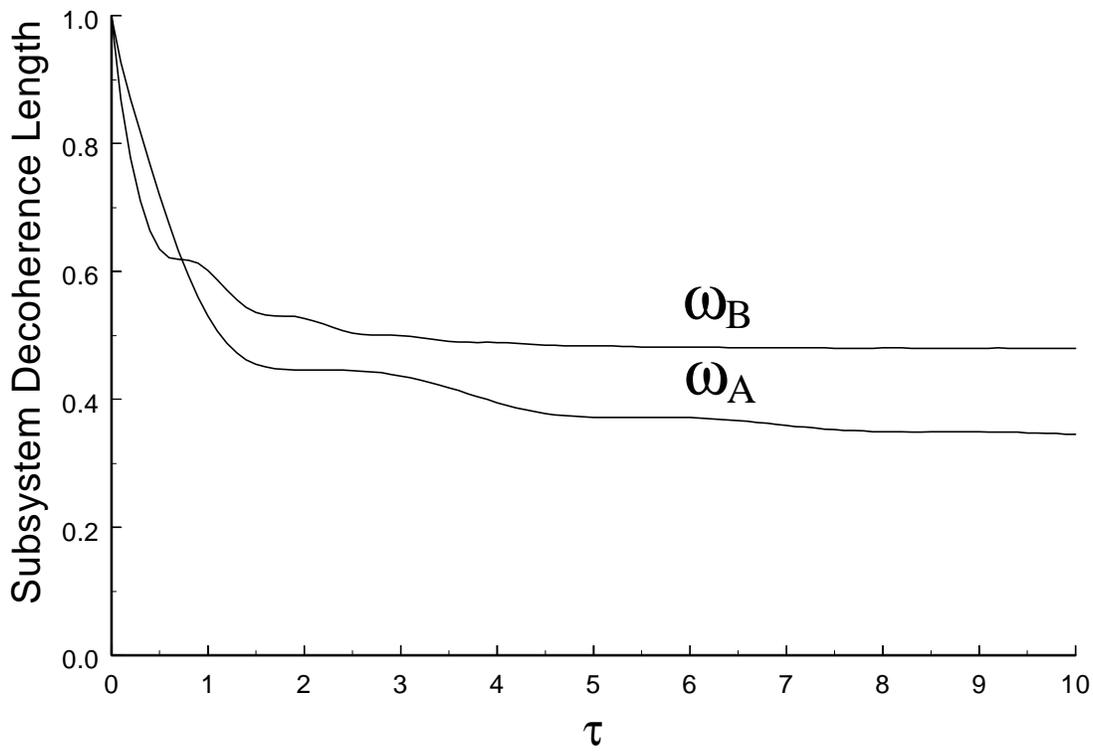

Figure 2. Subsystem decoherence lengths with the same parameters as in Figure 1.

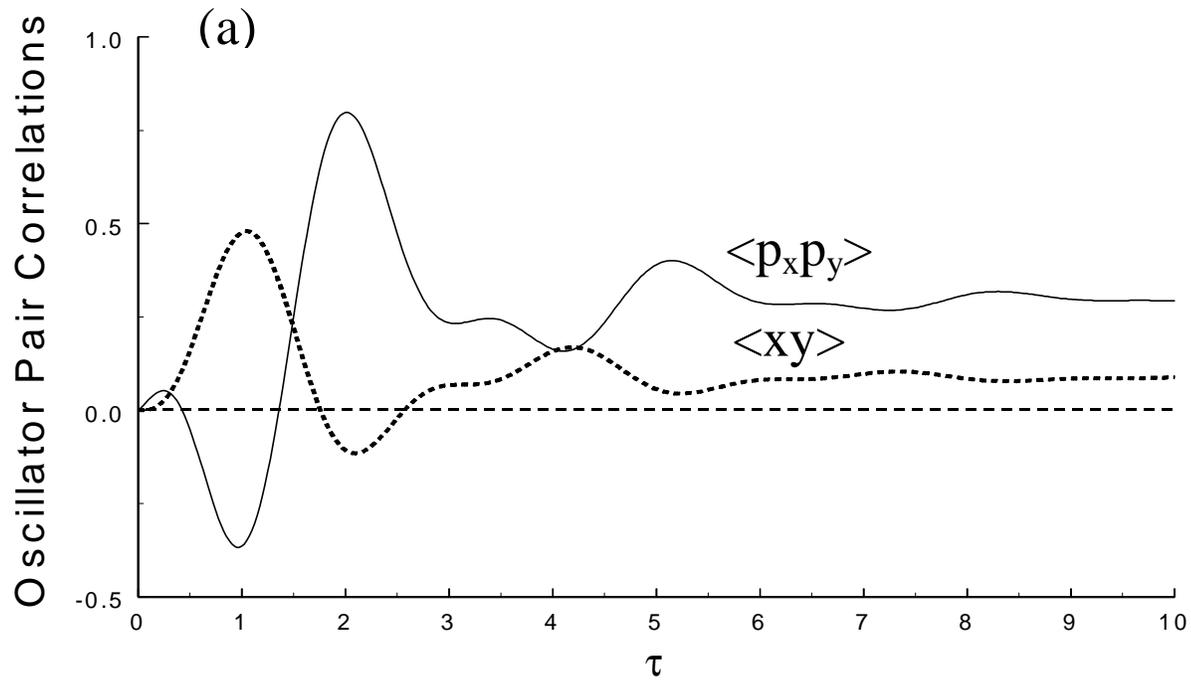

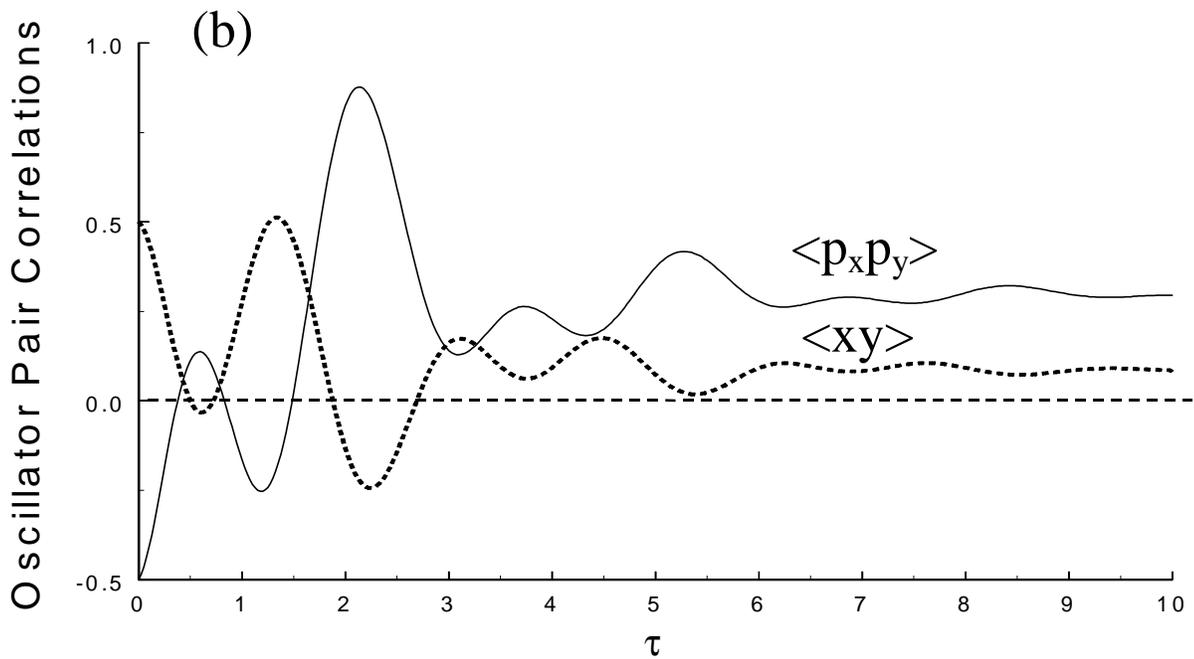

Figure 3. Oscillator pair correlations with the same parameters as in Figure 1 and (a) $a_{12} = 0 = b_{12}$ (unentangled initial state); (b) $a_{12} = 0.5 = -b_{12}$ (entangled initial state).

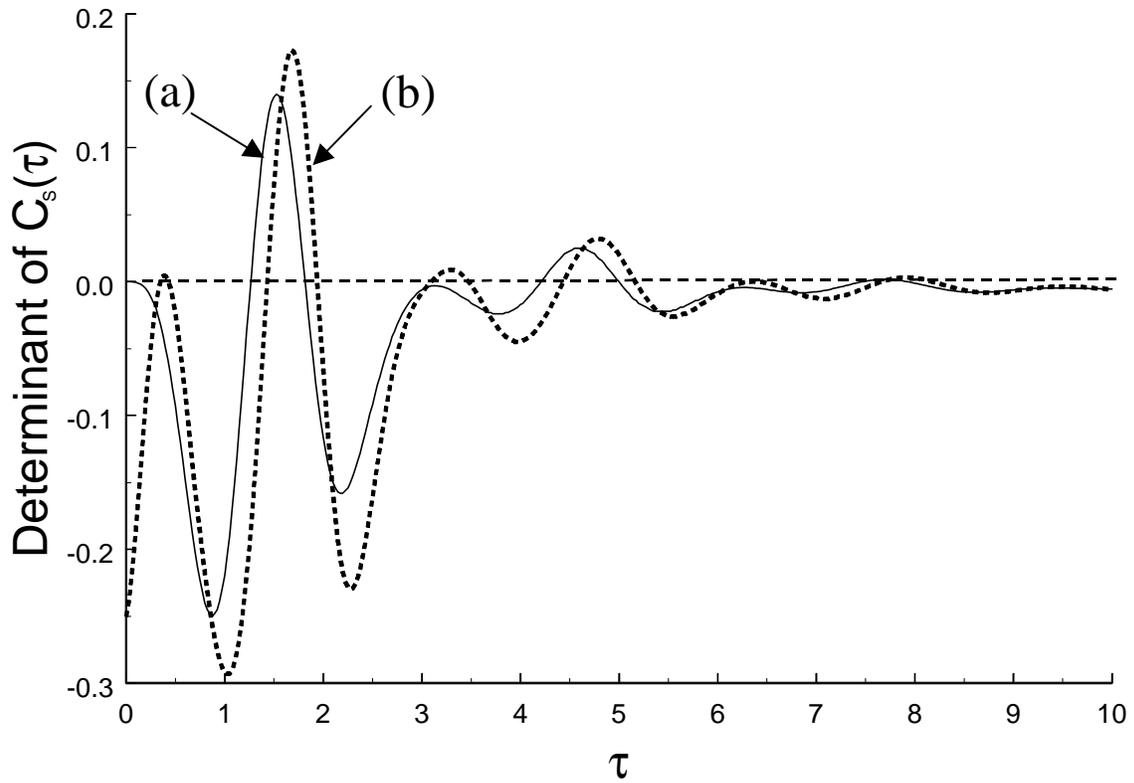

Figure 4. Evolution of the determinant of the covariance matrix $C_S$ (Eq.(40), the sign of which determines whether the state is entangled (<0) or unentangled (≥0), with the same parameters as in Figure 1 and (a) $a_{12} = 0 = b_{12}$ (unentangled initial state); (b) $a_{12} = 0.5 = -b_{12}$ (entangled initial state).